\documentclass[aps,prl,twocolumn,superscriptaddress,altaffillsymbol]{revtex4-2}

\usepackage[autostyle=true]{csquotes}

\usepackage{amsmath}
\usepackage{amsfonts}
\usepackage{mathtools}
\usepackage{amsbsy}
\usepackage[normalem]{ulem}
\usepackage{color}
\usepackage{graphicx}
\usepackage{dcolumn}
\usepackage{bm}
\usepackage{mathtools}
\usepackage{siunitx}

\usepackage[unicode=true,
bookmarks=true,bookmarksnumbered=true,bookmarksopen=true,bookmarksopenlevel=2,
breaklinks=false,pdfborder={0 0 1},backref=false,colorlinks=true,allcolors=blue]{hyperref}

\begin{document}

\author{Minghao Li}
\affiliation{Universit\'e de Strasbourg, CNRS, Institut de Science et d'Ing\'enierie Supramol\'eculaires, UMR 7006, 67000 Strasbourg, France.}
\author{Oussama Sentissi}
\affiliation{Universit\'e de Strasbourg, CNRS, Institut de Science et d'Ing\'enierie Supramol\'eculaires, UMR 7006, 67000 Strasbourg, France.}
\author{Stefano Azzini}
\altaffiliation[Present address: ]{Nanoscience Laboratory, Physics department, University of Trento, I-38123, Povo, Trento, Italy.}
\affiliation{Universit\'e de Strasbourg, CNRS, Institut de Science et d'Ing\'enierie Supramol\'eculaires, UMR 7006, 67000 Strasbourg, France.}
\author{Cyriaque Genet}
\email[]{genet@unistra.fr}
\affiliation{Universit\'e de Strasbourg, CNRS, Institut de Science et d'Ing\'enierie Supramol\'eculaires, UMR 7006, 67000 Strasbourg, France.}

\title{Galilean relativity for Brownian dynamics and energetics}

\begin{abstract}
We study experimentally the impact of inertial reference frame changes on overdamped Brownian motion. The reference frame changes are implemented by inducing, with a laser, laminar convection flows in a column of fluid where Brownian microspheres are dispersed. The convection flow plays the role of the relative velocity between the laboratory and the fluid comoving frames, and enables us to analyse the consequences of Galilean transformations on Brownian diffusion. We verify in particular how the Brownian dynamics remains ``weakly'' Galilean invariant, in agreement with recent discussions \cite{CairoliPNAS2018}. We also carefully look at the consequences of Galilean relativity on the Brownian energetics. This leads us to derive a Galilean invariant expression of the stochastic thermodynamic first law, consistent with existing theoretical results \cite{SpeckPRL2008}. We finally discuss a potential ambiguity of the Galilean relativistic features of diffusive systems that has obvious practical implications in the context of force measurements in external flows. 
\end{abstract}

\maketitle

%%%%%%%%%%%
\section{Introduction} 

The laws of classical mechanics are written in inertial reference frames, interconnected by Galilean transformations (GT) that ensure the invariance of the acceleration. If one further assumes that masses and forces are invariant through GT, Newton's laws then become invariant within the whole class of inertial reference frames, according to the principle of Galilean relativity \cite{RindlerBook}. 

It is however well-known that stochastic diffusive models break Galilean invariance (GI) because friction and random forces entering the stochastic equations of motion, such as the Langevin equation, emerge from coarse-graining procedures performed in the preferred reference frame where the fluid is at rest \cite{Kubo1986,ZwanzigJSP1973}. This selection of a reference frame is in contradiction with GI and this difficulty demands proper transformation rules for describing the coarse-grained dynamics in different inertial reference frames \cite{ZwanzigJSP1973}. Such rules have been recently given theoretically, from both dynamic \cite{CairoliPNAS2018} and energetic \cite{SpeckPRL2008} view points. The Langevin description of the Brownian motion between the preferred and a moving inertial frames was shown to differ only by a drift term that corresponds to the difference in the thermal noise statistics between the two frames. With this difference, the Langevin equations written in each frame are related by GT performed on positions and velocities only. This led in particular to maintain a ``weak'' GI for the description of the stochastic system with motional probability density functions (PDF) that are only shifted according to the GT that interconnects the two frames \cite{CairoliPNAS2018}. A fundamental consequence in this relativistic framework is the necessity to modify the stochastic energetics in order to account for this drift term with frame invariant definitions of work, heat and entropy (see Appendix C) \cite{SpeckPRL2008}. As a consequence, it is crucial to carefully recognize and evaluate flows in Brownian experiments, such as Brownian dynamic force measurements using the drift model \cite{Bechinger2011force}, Brownian heat engines \cite{albay2021shift, holubec2020active}, Brownian systems with growing domains \cite{le2020continuous}, etc.

\begin{figure}[htb]
  \centering{
    \includegraphics[width=0.6\linewidth]{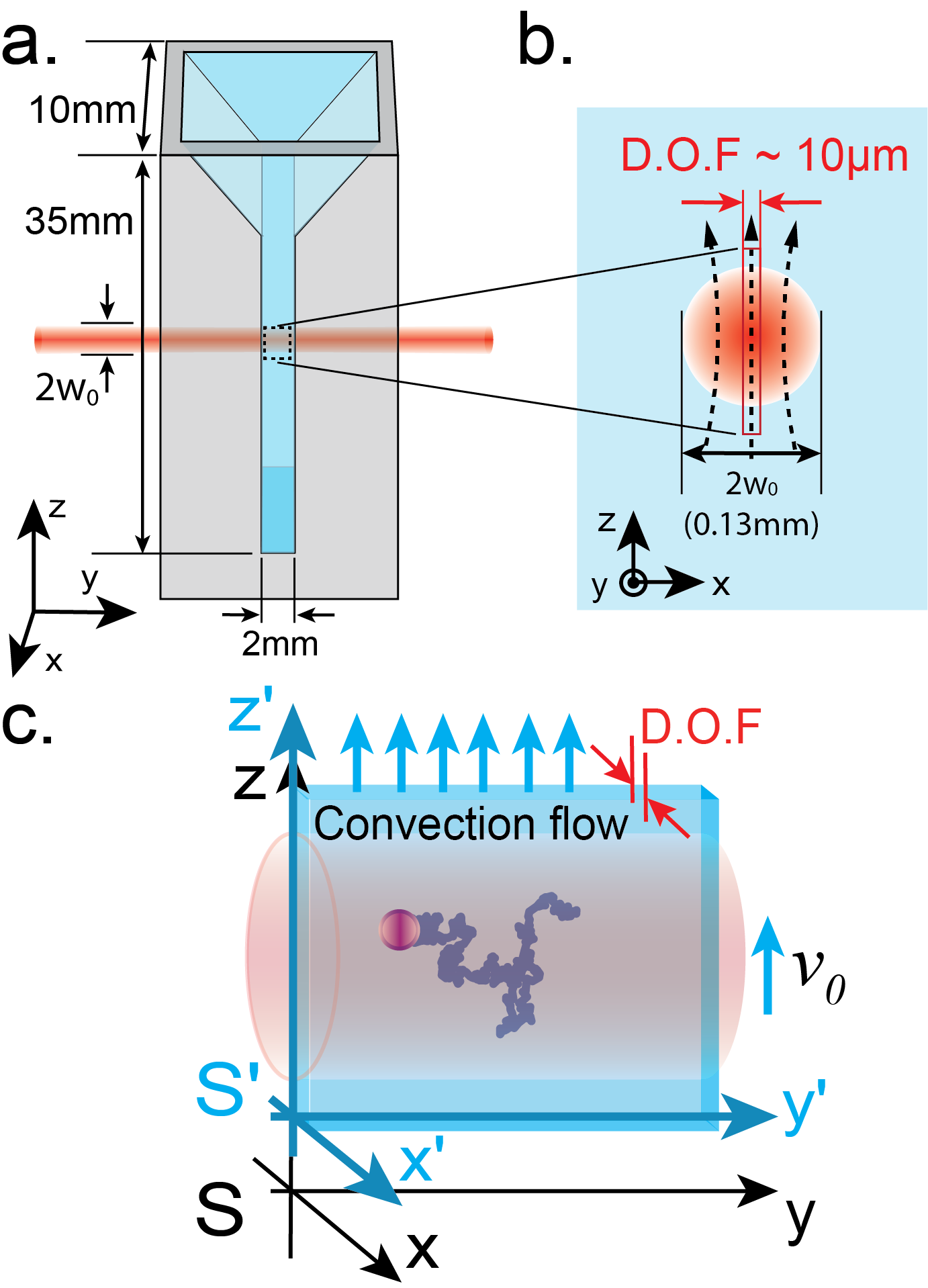}}
    \caption{Schematics of the experiment. (a) The cell containing the colloidal suspension of melamine microspheres in water consists of a quartz cuvette of dimensions $10(x)\times 2(y)\times 35(z) $ mm$^3$. The counterpropagating laser beams are sent through the centre of the cuvette along the $y-$axis. The region of interest (ROI) is chosen at the centre of the cell far from any wall, allowing us to neglect any perturbation of the walls on the diffusion dynamics. (b) The profile of the laser beam with a waist of $65~\mu$m is viewed in the $(x,z)-$plane where the arrows indicate qualitatively the convection flows due to the laser heating water. The depth-of-field determining the dimension of the ROI in the $x-$direction is of ca. $10~\mu$m within which the convection can be considered as laminar and uniform. (c) This laminar convection flow within the ROI act as a Galilean transformation for the Brownian diffusion along the $z-$axis, interconnecting, with a velocity ${v}_0 \hat{z}$, the two inertial reference frames, the laboratory frame $S$ and the comoving, fluid rest frame $S^\prime$. }
\label{fig1}
\end{figure}

In this Article, we experimentally explore these issues, by measuring the GT rules on an overdamped diffusive system under a well controlled laminar flow and by assessing the dynamic and energetic consequences of Galilean relativity on Brownian motion. We extend the discussion to the definition of Galilean invariant energetics quantities (work, heat, potential energy, entropy) and to the evaluation of the different PDF associated with the first law. These PDF are built experimentally and their transformations through GT are verified. The analysis reveals the crucial importance in distinguishing the Galilean relativistic drift from an `induced' force field in order to correctly describe the thermodynamics features of an overdamped Brownian system in a comoving frame. In addition, the control obtained on GT through the stable laminar flow allows a fine-tuning of colloidal levitating regimes that can be exploited in optofluidic systems for colloidal transport \cite{chen2020TOF, Quidant2021long}. This tuning is also potentially interesting to implement when studying the mechanical response of colloidal ensembles under weak external force fields \cite{canaguier2013mechanical, rukhlenko2016, LiPRA2019} or hydrodynamic interactions in the context of ordering effects and phase transitions in colloidal assemblies \cite{agarwal2009low,  janai2016dipolar}.

\section{Experimental inertial frame changes}

Our experiment consists in recording under laser illumination the Brownian motion of melamine micron-sized, spherical, beads dispersed in water inside a fluidic cell \cite{LiPRA2019}. The experimental configuration is described in Fig. \ref{fig1} and further in Appendix A. When the laser is off, the beads simply diffuse and sediment in the laboratory frame along the vertical axis. But as soon as the laser is switched on, water inside the cell heats up, with local modifications of its density $\rho$. Under conditions detailed below, this laser heating effect brings the fluid into uniform motion that can be precisely controlled along the vertical axis. The beads, dragged by the constant hydrodynamic flow hence generated, diffuse in a reference frame comoving with the fluid and interconnected to the laboratory frame through a GT. We monitor in the laboratory frame the Brownian motion of the beads and describe the dynamic and the energetic features associated with the GT.

Technically, specific requirements have to be met that determine the configuration schematized in Fig. \ref{fig1}. First, the fluidic cell is chosen sufficiently large so that it is possible to define in its centre an imaging Region of Interest (ROI) far from the cell walls so that the hydrodynamics within the ROI is described without the influence of boundary-wall conditions. The cell therefore is traversed by collinear, counterpropagating Gaussian laser beams of common waist $w_0\simeq 65~\mu$m with a Rayleigh length $z_R=18$ mm much larger than the cell width. Within the ROI, the Gaussian profiles of the laser beams can be considered as uniform along the optical $y-$axis. The waist is also much larger than the diameter of a single bead so that large statistical ensembles of displacements can be measured within the ROI. The dimension of the ROI along the observation $x-$axis is set by the depth-of-field (DOF) of the imaging objective (in our case, ca. $10~\mu$m) which is smaller than the laser waist. Finally, the small volume fraction $\phi \sim 10^{-6}$ of the micron-sized colloidal dispersion used in the experiment is such that the Brownian motion monitored within the ROI can be described in the absence of any hydrodynamic interaction between the beads. The practically plane-wave illumination conditions minimize any gradient contribution in the optical force field thus only determined by scattering contributions, i.e. radiation pressure. The counterpropagating beam configuration allows to cancel any effect of radiation pressure \cite{LiPRA2019} by simply tuning the intensities in each beams to even values. 

With such dimensions and conditions of illumination, the laser sets the fluid into laminar motion inside the ROI by a heat convection effect. Such a convective dynamics is described by coupling the equation of heat under laser illumination to the Navier-Stokes (NS) equation for the transport of fluid momentum per unit volume $\rho {\bf{v}}$ in the presence of a diluted, homogeneous, colloidal dispersion. Within the Boussinesq's approximation  \cite{Guyon2015hydro}, the change in density associated with the laser heating of the fluid $\delta\rho = -\alpha\rho\delta T$ is assumed to be such that $\delta\rho \ll \rho$, where $\alpha$ is the thermal expansion coefficient of water and $\delta T=T({\bf r})-T_0$ the difference between the local temperature and the background temperature of water inside the cell. 

Under such an approximation, the heat and NS coupled equations write as: 
\begin{eqnarray}
&&\rho c_p D_t \delta T = k\nabla^2 \delta T + \dot{q}_L  \label{Eqcoup1}\\
&&\rho D_t {\bf v} = \mu \nabla^2 {\bf v} -\alpha\rho\delta T {\bf g}+\phi\Delta\rho {\bf g} \label{Eqcoup2}
\end{eqnarray} 
with $c_p$ the specific heat capacity of water, $k$ its thermal conductivity, $\mu$ its shear viscosity --we assume standard (room temperature) values $c_p=4.18\times 10^3$ J/K/kg, $k=0.62$ W/K/m and $\mu=0.87\times 10^{-3}$ Pa s. On the heat transport equation, $\dot{q}_L = 2A P_0/\pi w_0^2 \exp{(-2(x^2+z^2)/w_0^2)}$ is the volumetric heat rate generated, at its waist, by the Gaussian laser of power $P_0$, taking for water an absorption coefficient $A=0.3$/m at a wavelength of $633$ nm. On the NS equation, $\phi\Delta\rho {\bf g}$ corresponds to the external body force exerted on a unit volume of water by the sedimenting ensemble of colloidal spheres with ${\bf g}=-g\hat{z}$ the gravitational acceleration, $\Delta\rho$ the density difference (513 kg/m$^{3}$) between a single melamine sphere and water and $\phi$ the volume fraction inside the ROI ($\phi\sim 10^{-6}$). 

\begin{figure}[htb]
  \centering{
    \includegraphics[width=0.8\linewidth]{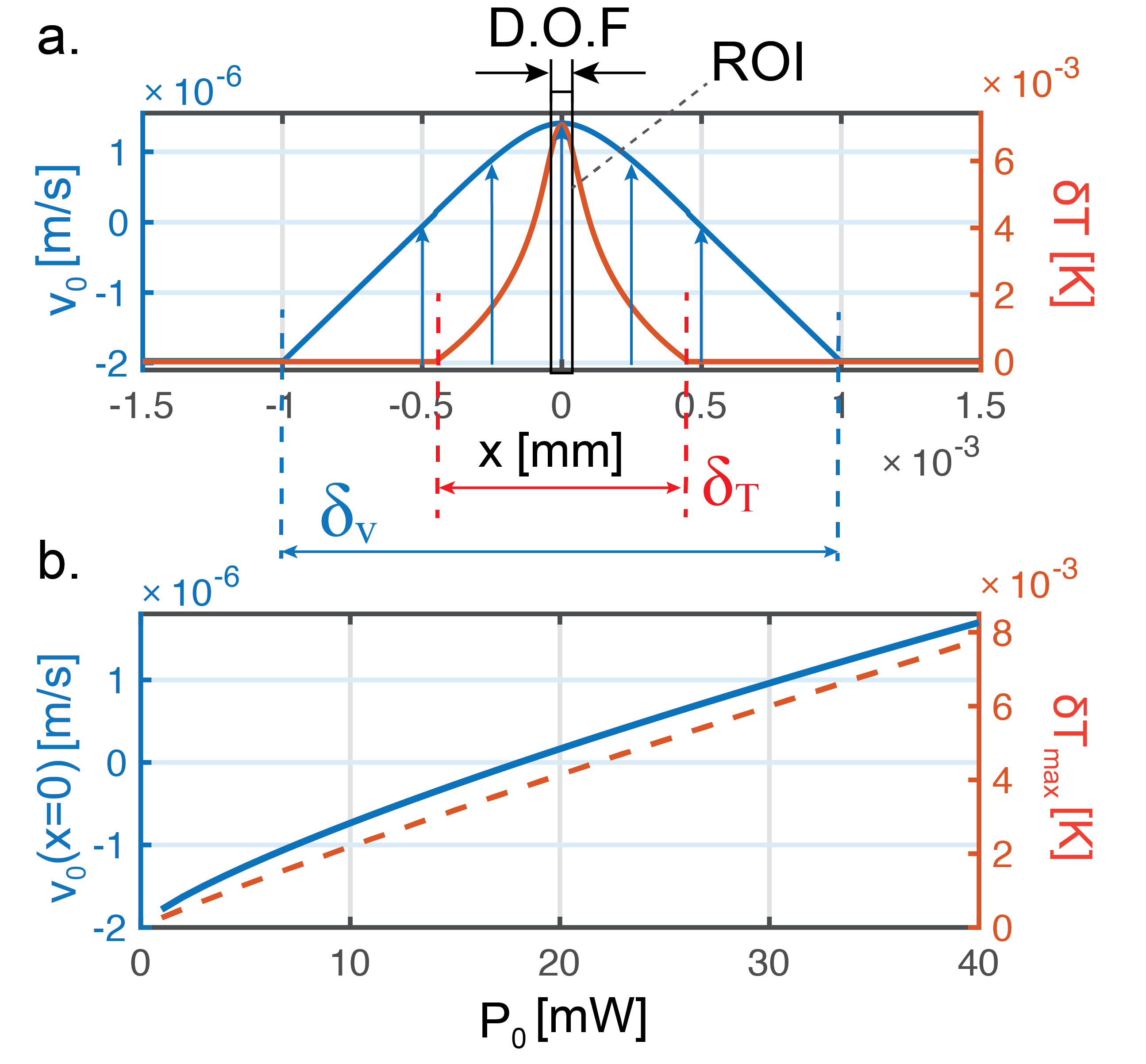}}
    \caption{(a) Laser-induced temperature difference $\delta T$ and frame change velocity $v_0$ profiles due to the laser-induced convection, calculated within the ROI and along the $x-$axis by the approximated solution of Eqs. (\ref{Eqcoup1},\ref{Eqcoup2}) detailed in appendix B for a laser power of $36$ mW and a mean volume fraction $\phi \sim 6.8 \times 10^{-7}$ which corresponds the order of magnitude of our experimental conditions. Temperature and velocity boundary layer thicknesses are marked as $\delta_T$ and $\delta_v$, respectively. (b) Profile maxima $\delta T_{\rm max}$ and frame change velocity within our ROI $v_{0}(x = 0)$ as a function of the illumination laser power. }
\label{fig2}
\end{figure}

Because of the very thin fluid layer defined within the imaging DOF thickness $\ell_{\rm DOF}\sim10~\mu$m --see Fig. \ref{fig1} (b)-- the velocity ${\bf v}({\bf r})$ of the convection flow is such that $v_x\sim 0$. We also assume $v_y \sim 0$ within the ROI positioned far from all walls. This is fully consistent with our observations that reveal a convection flow laminar within the DOF layer in the vertical $z-$direction, yielding therefore ${\bf v}({\bf r})=v_0({\bf r})\hat{z}$ with $v_0({\bf r})$ corresponding to the velocity of frame change. With the Boussinesq's approximation that preserves the incompressibility condition $\nabla\cdot {\bf v}=0$, we further write ${\bf v}({\bf r})=v_0(x,y)\hat{z}$. This cancels the convective contribution in the Lagrangian derivative where $D_t {\bf v} \sim \partial_t {\bf v}$ in the NS equation. Since the measured convection flow are of the order of $10^{-6}$ m/s and temperature changes $\delta T$ under ca. $100$ mW laser irradiation that we evaluate to be at the mK level, the convective contribution to $D_t \delta T$ can also be neglected in the heat equation. As a consequence, the two equations are decoupled and can be solved in the steady-state, as detailed in Appendix \ref{AppCnvt}. The corresponding thermal and velocity profiles are plotted in Fig. \ref{fig2}. Temperature and velocity boundary layer are evaluated as $\delta_{\rm T} \sim 0.9$ mm and $\delta_{\bf v} \sim 2$ mm respectively. With $\delta_{\bf v} \gg \ell_{\rm DOF}$, the convection flow is strictly laminar within the DOF and uniform across the ROI. The central point of our scheme is that in the steady-state regime, this flow uniformly carries the colloidal beads and hence defines an inertial reference frame $(z',t')$ where the fluid is at rest. This comoving frame $S^\prime$ is related to the laboratory frame $S$ by a GT where the velocity of $S^\prime$ with respect to $S$ is set and controlled by the laser illumination power $P_0$. 

\section{Brownian dynamics in different inertial frames}

In the comoving inertial frame $S^\prime$, the Brownian motion of each bead is performed under the constant sedimentation force field $\Delta\rho V {\bf g}$ resulting from the competition between gravity and buoyancy. The motion is described by the Langevin equation written along the $z'-$axis as:
\begin{equation}
\gamma \dot{z'} = -\Delta \rho Vg + \sqrt{2k_BT\gamma}\cdot\xi'(t'),
\label{Langfl}
\end{equation} 
where {$\xi'$ is the random thermal force with $\left< \xi'(t') \right> = 0$ and $\left< \xi'(t'_1) \xi'(t'_2) \right> = \delta(t'_1-t'_2)$, $\gamma$ is the friction coefficient, $k_B$ is the Boltzmann constant and $T$ is the temperature. 

Moving to the laboratory frame $(z,t)$ can be simply done by the GT $z=\mathcal{G}[z']=z'+v_0t$, $t=\mathcal{G}[t']=t'$ performed on the velocity of the Langevin equation (\ref{Langfl}) while leaving the noise $\xi'$ unchanged to give:
\begin{equation}
\gamma \dot z = -\Delta \rho Vg + \sqrt{2k_BT\gamma}\cdot\xi'(t) + \gamma v_0.
\label{Langlab}
\end{equation} 

As explained in \cite{CairoliPNAS2018}, the possibility to do so is physically rooted in the fact that the drift term $\gamma v_0$ induced by the GT fundamentally corresponds to the modification of the thermal noise statistics between the two inertial frames, with $\sqrt{2k_{\rm B}T\gamma}\xi' (t')+\gamma v_0 = \sqrt{2k_{\rm B}T\gamma}\xi (t)$. This connection leads to a ``weak'' GI of the Langevin equation. This additional drift has important thermodynamic consequences that we discuss below.

\begin{figure}[htb]
  \centering{}
    \includegraphics[width=1\linewidth]{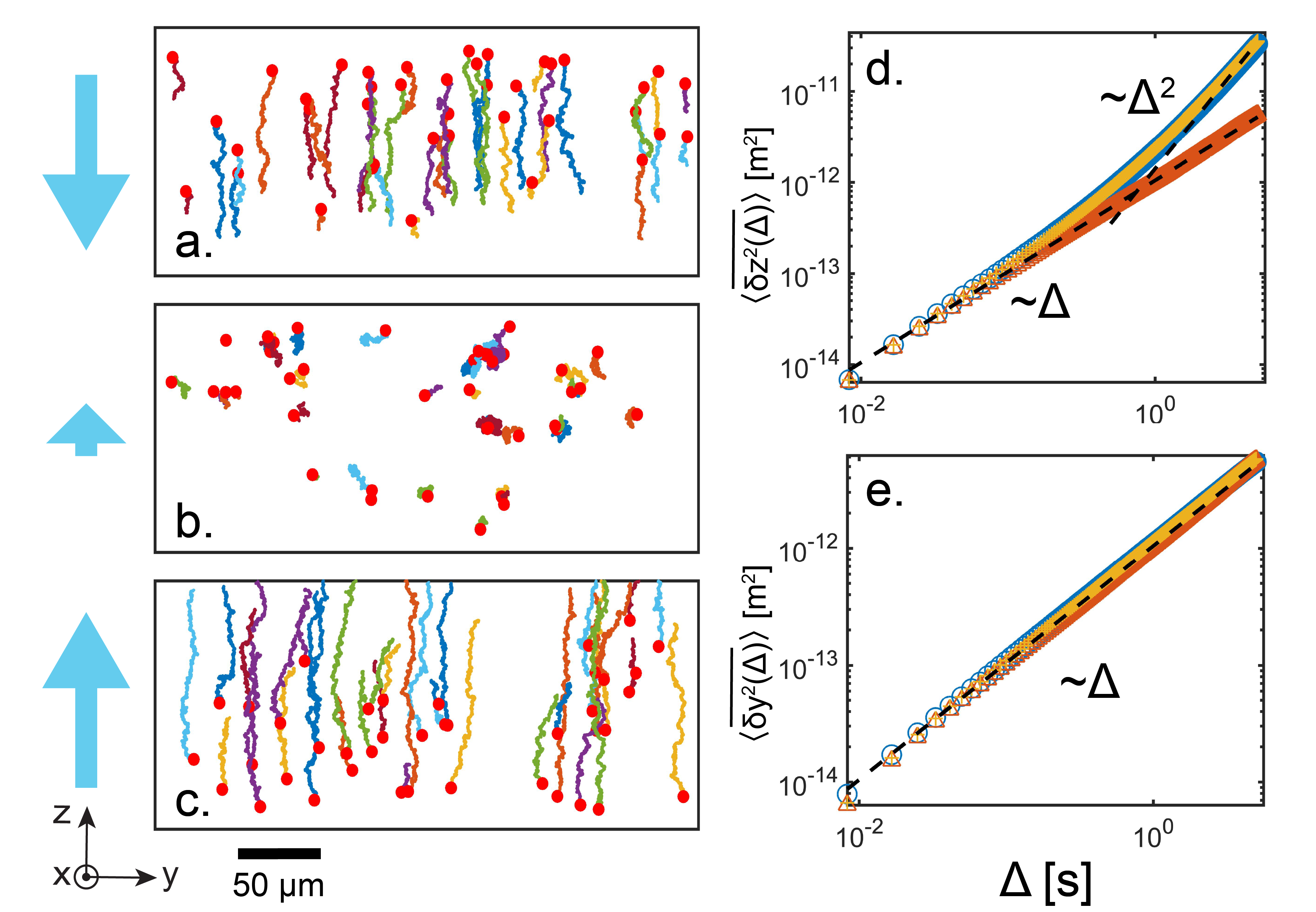}
    \caption{(a)-(c). Individual experimental trajectories recorded in the $(y,z)$ plane of the ROI for 3 different regimes controlled by the laser power: collective sedimentation regime (low laser power at 10 mW), suspension regime (laser power tuned at 24 mW) and lift regime (high laser power at 36 mW), with the starting point of each trajectory marked by a red dot. The global flow direction and its relative strength are indicated by the blue arrows on the left-hand side of the graphs. (d)-(e) Mean Square Displacements (MSD)  calculated in both the $z$ --panel (d)-- and $y$ --panel (e)-- directions from the trajectories observed in the 3 regimes presented in (a) -blue circles, (b) -red triangles- and (c) -orange crosses. The MSD along the $y$ direction in the 3 regimes remains linear, revealing a normal Brownian diffusion with no external flow or force acting on this direction. In contrast, the MSD along the $z$ direction in regimes (a) and (c) are parabolic. But for a well-tuned laser power, the MSD along the $z$ direction can remain linear, corresponding to the remarkable suspension regime observed in the laboratory frame on the trajectories displayed in (b).}
\label{fig3}
\end{figure}

We first analyze the Brownian motion in the laboratory frame under different laser illumination powers. All the corresponding trajectories of the microspheres within the ROI are recorded from successive images that give access to the succession of vertical displacements $\Delta z_i(t_k ) = z_i (t_{k+1} ) - z_i (t_k )$ measured, for one trajectory $i$, at a fixed frame rate $f=1/(t_{k+1}-t_{k})$. When the laser power is weak, the microspheres collectively sediment in the cell with an ensemble of trajectories displayed in Fig. \ref{fig3} (a). By increasing the laser power, the convection flow is induced and drags the spheres upward. It is easy to find a power value (i.e. a convection velocity) that can practically compensate sedimentation, leaving the spheres suspended in the laboratory frame. The trajectories corresponding to this case are shown in Fig. \ref{fig3} (b). Convection can even take over sedimentation if the laser power is further increased, as seen clearly in Fig. \ref{fig3} (c). These trajectories can be analyzed, as a function of the time lag $\Delta$, by the mean square displacement (MSD) averaged on the ensemble of $N$ trajectories recorded within the ROI
\begin{equation}
\langle \delta ^2 z(\Delta)\rangle =\frac{1}{N}\sum_i \left[ z_i (t+\Delta)-z_i(t)\right]^2 ,
\label{ensMSD}
\end{equation}
a stationary quantity independent of the initial time $t$. The three MSD plotted on Fig. \ref{fig3} (d) clearly reveal the dynamics in the three different cases, with parabolic $\Delta ^2$ MSD in both sedimentation and convection regimes. For the suspended case, the MSD is practically linear in $\Delta$, a feature that corresponds to the free-like Brownian dynamics observed in Fig. \ref{fig3} (b). We also show in Fig. \ref{fig3} (e) that the MSD evaluated from the ensemble of displacements recorded along the $y-$axis remains perfectly linear, confirming that the convection flow is only induced by the laser along the vertical $z-$axis, implementing the GT detailed above.

From these ensembles, it is also possible to construct the displacement probability density functions (PDF) associated with each regime of convection. In the comoving fluid frame, the Brownian spheres diffuse in the gravity force field characterized by a sedimentation velocity ${v}_{\rm sed}=-\Delta\rho V {g}/\gamma$ along the $z'-$axis according to Eq. (\ref{Langfl}). In this frame therefore, the PDF for vertical displacements is given by $P'(\Delta z',\Delta t')=\exp{(-(\Delta z'-{v}_{\rm sed}\Delta t')^2/4 D\Delta t')}/\sqrt{2\pi D \Delta t'}$. Here, $\Delta z'$ is a displacement measured along the $z' -$axis within a time lag $\Delta t'$, and $D=k_{\rm B}T/\gamma$ the diffusion coefficient in the vertical direction. In the laboratory frame, Eq. (\ref{Langlab}) yields 
\begin{equation}
P(\Delta {z},\Delta {t}) = \frac{1}{\sqrt{2\pi D \Delta {t}}}\exp\left(- \frac{\left(\Delta{z}- v_z \Delta{t} \right)^2}{4D \Delta {t}} \right)  \label{PDFLab}
\end{equation}
where $v_z=v_0+v_{\rm sed}$ is the mean velocity along the $z-$axis in the laboratory frame, resulting from sedimentation and convection flows. The comparison between $P'(\Delta z',\Delta t')$ and $P(\Delta z,\Delta t)$ verifies the relation between the PDF acquired in different inertial frames --see below \cite{CairoliPNAS2018}:
\begin{equation}
P(z, t) = P^{\prime}(z-v_0 t, t).
\label{Ptrans}
\end{equation}

The experimental PDF constructed in the laboratory frame are displayed in Fig. \ref{fig4} (a-c) for the three different values of laser illumination power. We also extract from these PDF the mean velocity $v_z = \left< \Delta z/ \Delta t \right>$ whose evolution as a function of the laser power is plotted in Fig. \ref{fig4} (d). The dispersions in the $v_z$ values measured for fixed power levels show that $v_z$ is relatively stable in time when averaged throughout the ROI, confirming that the combination of sedimentation and convection flows describes a GT. Since the PDF plotted in the laboratory frame mix displacements measured on different trajectories and at different times, it is interesting to extract the diffusion coefficient $D$ from a fit of the PDF variances at each time lag $\Delta t$. All $D$ thus extracted are plotted in Fig. \ref{fig4} (e) for all different laser powers. The agreement between these values extracted from experimental data and the theoretically expected diffusion coefficient (including uncertainties associated with the size dispersion of the colloidal suspension) demonstrates that the laser induced convection flow corresponds to a genuine drift term and does not affect the Brownian noise spectrum. This constitutes an experimental proof that a change of inertial reference frames modifies the noise spectrum as $\xi(t)= \xi' (t')+v_0 \gamma / \sqrt{2k_{\rm B}T\gamma}$. As discussed above, this drift term exactly corresponds to the GT performed directly on the velocity $\dot{z'}$, preserving the fluctuation-dissipation theorem \cite{kubo1966fluctuation}.

\begin{figure}[htb]
  \centering{}
    \includegraphics[width=0.8\linewidth]{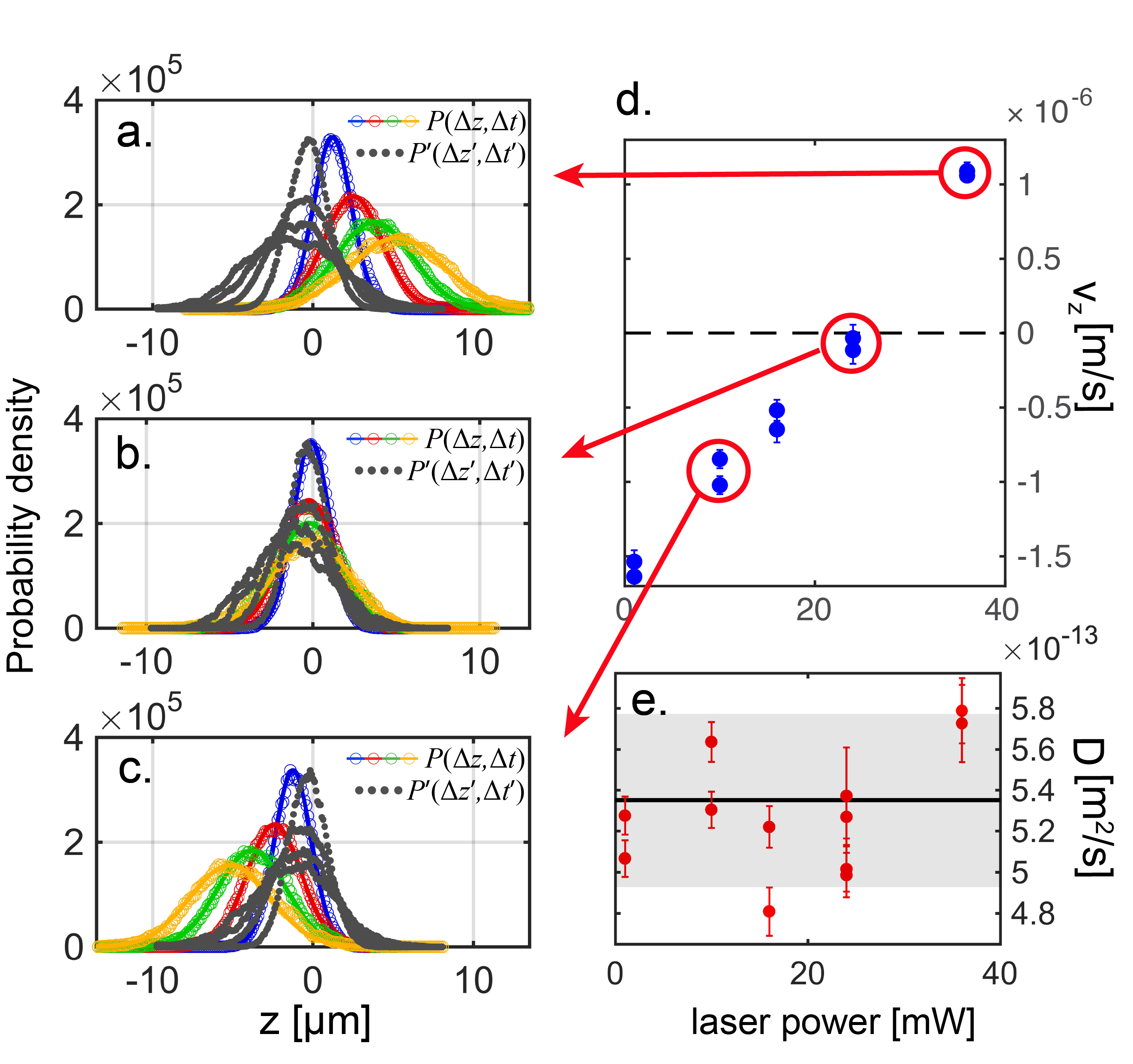}
    \caption{(a)-(c) Probability density functions (PDF) for the displacements measured along the $z-$axis in the laboratory frame $P(\Delta z,\Delta t)$ evolving with time lag $\Delta t$ from $1$ s to $4$ s --colored data points-- and PDF for the vertical displacements of Galilean transformed reconstructed trajectories (see main text) shown as grey data points evolving with the time lag $\Delta t$ from $1$ s to $4$ s. Such a PDF corresponds to the dynamics expected in the fluid rest frame ($S^\prime$) $P^\prime(\Delta z^\prime, \Delta t^\prime)$. One clearly sees that all the PDF hence reconstructed are all similar, while they correspond each to different dynamical regimes when observed from the laboratory frame. These regimes are measured for the different particle velocities $v_z$ shown in (d) for the corresponding laser powers. These velocities $v_z=\langle \Delta z/\Delta t\rangle$ are measured by time ensemble averages of vertical displacements $\Delta z$ in the laboratory frame --within the smallest time lag $\Delta t = 1/120$ corresponding to the inverse of frame rate-- divided by $\Delta t$ with the standard error indicated in the data as an error bar. Positive values for $v_z$ correspond to convection flows moving upwards. When $v_z<0$, the convection flow is not strong enough (low laser power regimes) to compensate the collective sedimentation flow. (e) Brownian diffusion coefficients along the $z-$axis (red points) obtained by fitting linearly the variances of the PDF for each $\Delta t$ with error bars coming from the $99\%$ linear fitting confidence interval. Each variance is obtained by a Gaussian fit made on each PDF. The value of the diffusion coefficient expected from the measurement under the condition of temperature ($T = 299$ K), viscosity ($\eta = 0.87\times10^{-3}$ N.s.m$^{-2}$) and particle diameter ($d = 0.94 ~\mu$m) is also displayed (black solid line) with the inaccuracy coming from the remaining uncertainty in the temperature determination ($\delta T = \pm 1$ K), viscosity ($\delta\eta = \pm 0.02\times10^{-3}$ N.s.m$^{-2}$) and particle diameter ($\delta d = \pm 50$ nm). This inaccuracy in the determination of the diffusion coefficient is represented by the grey shaded zone.}
\label{fig4}
\end{figure}

With such a flow stable in time and homogeneous in space within the ROI, the comoving frame velocity $v_0$ can be estimated from the determination of $v_z$ using the PDF $ P(\Delta {z},\Delta {t})$ and the knowledge of $v_{\rm sed}$ assuming that the sole external force field exerted along the $z-$axis results from buoyancy. With this, the trajectories in the comoving fluid frame can be reconstructed by applying a GT to each trajectory recorded in the laboratory frame. The PDF in this frame can then be built and they are plotted in panels (a-c). The comparison clearly shows that the laboratory PDF are related to the comoving PDF by a GT, $\Delta {z}=\mathcal{G}[\Delta z']=\Delta z'+v_0\Delta t'$ with $P(\Delta {z},\Delta {t})=\mathcal{G}[P'(\Delta z',\Delta t')]=P'(\mathcal{G}[\Delta z'],\Delta t)$ with $P'(\Delta z',\Delta t')$ defined above. This is in perfect agreement with the ``weak'' GI principle proposed by \cite{CairoliPNAS2018}.

\section{Galilean invariance for Brownian stochastic energetics}

We now look carefully at the energetic consequences of such a change of reference frame. To do so, the influence of the convection flow must be accounted for in the definition of the production rates associated with the stochastic thermodynamics of Brownian motion, while keeping in mind that the drift term related to GT fixes the choice of the reference frame.

Following Sekimoto's approach \cite{SekimotoBook},  the heat transferred to the fluid by the Brownian system is described, in the fluid reference frame, by a force $F'_b = -\gamma \dot{z'}+\sqrt{2k_{\rm B}T\gamma}\xi' (t')$ that the thermal bath (the fluid) exerts on the beads along the vertical axis. This force combines both the friction and the random force, and it exactly compensates the external force field acting on the system according to the Langevin equation. Most generally, this external force field writes as $F'(z',t')=-\nabla U'(z',t') + f'(z',t')$ with $f'$ a non-conservative force. In the fluid reference frame therefore, the heat production rate can be expressed as a function of the external force as \cite{SeifertReview2012}:
\begin{equation}
\dot{q'}=-F'_b\cdot \dot{z'}=F'(z',t')\cdot\dot{z'}.
\end{equation}

In the laboratory frame, the stochastic force $F_b$ written as $F_b = -\gamma \dot{z}+\sqrt{2k_{\rm B}T\gamma}\xi (t)$ will account for the difference in the noise statistics $\sqrt{2k_{\rm B}T\gamma}\xi' (t)+\gamma v_0 = \sqrt{2k_{\rm B}T\gamma}\xi (t)$ between both frames discussed above. Therefore, with the heat bath at equilibrium in the fluid rest frame, the force writes as
\begin{equation}
F_b = -\gamma(\dot{z}- v_0) + \sqrt{2k_{\rm B}T\gamma}\xi' (t),
\end{equation}
showing that the relevant displacement to be accounted for when describing heat exchanges between the sphere and the fluid thermal bath is the actual displacement $z-v_0 t$ of the Brownian sphere with respect to the fluid.

Just like for Newton's laws, Galilean relativity enforces $f(z,t)=f'(z-v_0t,t)$ for the non-conservative force, $U(z,t)=U'(z-v_0t,t)$ for the potential energy, and therefore $\nabla U(z,t)=\nabla U'(z-v_0t,t)$. This implies $F_b=-F(z,t)=-F'(z-v_0t,t)$ and the final expression of the heat production rate evaluated in the laboratory frame
\begin{eqnarray}
\dot{q} &=& F'(z-v_0t,t)\cdot (\dot{z}-v_0)  \nonumber  \\
&=& [-\nabla U'(z-v_0t,t)+f'(z-v_0t,t)]\cdot (\dot{z}-v_0).
\end{eqnarray} 
The heat production rate is clearly GI with $\dot{q}=\dot{q'}$ where $\dot{q}=\mathcal{G}[\dot{q'}]$.  Replacing the expressions of the potential energy and the non-conservative force in the comoving fluid frame by those in the laboratory frame yields $\dot{q}= [-\nabla U(z,t)+f(z,t)]\cdot (\dot{z}-v_0)$. 

With the same approach, we can write the rate of potential energy change as $\dot{u'}=dU'(z',t')/dt'=[\dot{z'}\cdot\nabla +\partial_{t'}]U'(z',t')$ in the comoving frame. In the laboratory frame, $\dot{u}=[\dot{z}\cdot\nabla +\partial_{t}]U(z,t)$ gives:
\begin{eqnarray}\nonumber
\dot{u}&=&\frac{dU'(z-v_0t,t)}{dt}\\
&=&[(\dot{z}-v_0)\cdot\nabla+\partial_t]U'(z-v_0t,t),
\end{eqnarray}
reminding that $U(z,t)=U'(z-v_0t,t)$. This shows that the rates are GI and connected from one reference frame to another by a GT performed on the position and velocity. That is $\dot{u}=\mathcal{G}[\dot{u'}]=\dot{u'}$.

This leads us to a GI description of the stochastic work production rate and therefore of the first law of thermodynamics. Indeed, in the fluid reference frame, according to the first law, $\dot{u'}=\dot{w'}-\dot{q'}$ so that $\dot{w'}=\partial_{t'}U'(z',t')+f'(z',t')\cdot z'$. In the laboratory inertial frame, $\dot{w}=\dot{u}+\dot{q}=\partial_t U'(z-v_0t,t)+f'(z-v_0t,t)\cdot (\dot{z}-v_0)$ shows that $\dot{w}=\mathcal{G}[\dot{w'}]$. The work production rate eventually can be written as
\begin{eqnarray}\nonumber
\dot{w}&=&[v_0\cdot\nabla+\partial_t]U(z,t)+f(z,t)\cdot (\dot{z}-v_0)  \\
&=&\dot{w'}
\end{eqnarray}
emphasizing the GI of the rate.

These general relations can be specified to our experimental case as soon as the relative velocity between the laboratory frame and the comoving fluid frame is measured. The force resulting from gravity and buoyancy $\Delta\rho V {\bf g}$ in Eq. (\ref{Langfl}), derives from a potential energy $U'(z',t')=\Delta\rho Vg(z'+v_0t')$ that determines the heat production rate $\dot{q'}=-\nabla U'(z',t')\cdot \dot{z'}=-\Delta \rho V g \dot{z'}$ in the fluid reference frame.  In the laboratory frame, ${U}({z},{t})= \Delta\rho Vg {z}=U'({z}-v_0 t,t)$, so that $\dot{q}=-\Delta \rho V g(\dot{{z}}-v_0)$ which precisely equals $\dot{q'}$ as expected from GI. This invariance corresponds to the fact that in the fluid reference frame, the stochastic heat production rate is independent of $v_0$ and only depends on the displacement $z$ measured in the comoving frame. 

In this comoving frame, the stochastic work production rate is given by the change in the potential energy as $\dot{w'}=\partial_{t'}U'(z',t')=\Delta\rho V g v_0$ which is entirely fixed by the convection velocity. In the laboratory frame, the change in the potential energy $U(z,t)$ has to account for the drag of the beads and thus involves the convective derivative $\dot{{w}}=\partial_t {U}({z},{t})+{v}_0\cdot\nabla {U}({z},{t})=\Delta\rho V g v_0$ \cite{SpeckPRL2008, GerloffPRE2018}. In this way, GI is ensured with $\dot{w'}=\dot{w}$. By this, we recover the expression for the rate of potential energy change using the first law $\dot{u'}=-\dot{q'}+\dot{w'}=\nabla U'(z',t')\cdot \dot{z'} + \partial_{t'} U'(z',t')=\Delta\rho V g (\dot{z'}+v_0)$ and its GI with $\dot{{u}}=\dot{u'}$.

The work production rate accounts for the \textit{deterministic} energy contribution injected by the illuminating laser in the system in order to put the whole column of fluid into motion, with a fixed heat production rate.  The work does not come from an external force exerted on the particle itself. As discussed further below, this distinction is important to appreciate in the energetic context. Our setup therefore yields a stochastic energetics different from the one at play on colloidal suspensions under shear flows (\cite{GerloffPRE2018},  e.g.) or when a Brownian particle is optically trapped and the trap is dragged through the fluid at a constant velocity. There, the fluid actually works on the confined particle, bringing it out of equilibrium. Such a scheme has been described in details in \cite{CohenJPS2008}.

The quantities determined experimentally from the displacement PDF are the PDF associated with these thermodynamic production rates. These PDF can be directly built from the motional PDF acquired in the chosen reference frame over a measurement time $\Delta t=1$ s for instance, giving us enough statistics with variances of the order of $k_{\rm B}T$. From Eq. (\ref{PDFLab}), we easily calculate:
\begin{eqnarray}
&&P(Q,\Delta t) = \frac{\exp {\left( - \frac{(Q + \Delta \rho Vg v_{\rm sed} \Delta t)^2}{4D\Delta t  (\Delta\rho Vg)^2}\right)}}{\Delta\rho V g \sqrt{2\pi D \Delta t}} \\
&&P(U,\Delta t) =  \frac{\exp {\left( - \frac{[U - \Delta\rho V g (v_0+v_{\rm sed})\Delta t]^2}{4D\Delta t  (\Delta\rho Vg)^2}\right)}}{\Delta\rho V g \sqrt{2\pi D \Delta t}}\\
&&P(W,\Delta t) =  \delta \left(W- \Delta\rho Vg v_0\Delta t\right) 
\end{eqnarray}
considering that $W$ has one value for a given convection velocity $v_0$. We set three different convection flows dragging the Brownian particles whose corresponding trajectories are displayed in Fig. \ref{fig5} (a-c) and the associated PDF shown in panels (d-f). With GI of the energetic quantities, the PDF are identical in all inertial frames. These PDF perfectly obey the first law with $\langle U \rangle=-\langle Q \rangle +\langle W \rangle$, knowing $v_{\rm sed} = \langle \dot z' \rangle$, $v_z = \langle \dot z \rangle$ and $v_{z} = v_{\rm sed} + v_{0}$. As seen on the data of Fig. \ref{fig5} (d) with a position of $P(W,\Delta t)$ in the negative energy scale, we start at the lowest laser power with a laser-induced convection not strong enough to compensate for the collective sedimentation effect of the body force. Increasing the laser power reverses the landscape, by passing through the remarkable regime already discussed above in Fig. \ref{fig3} (b), where the Brownian motion, seen in the laboratory frame, appears as practically free.

%% Figure 5 %%
\begin{figure}[htb]
  \centering{}
    \includegraphics[width=0.99\linewidth]{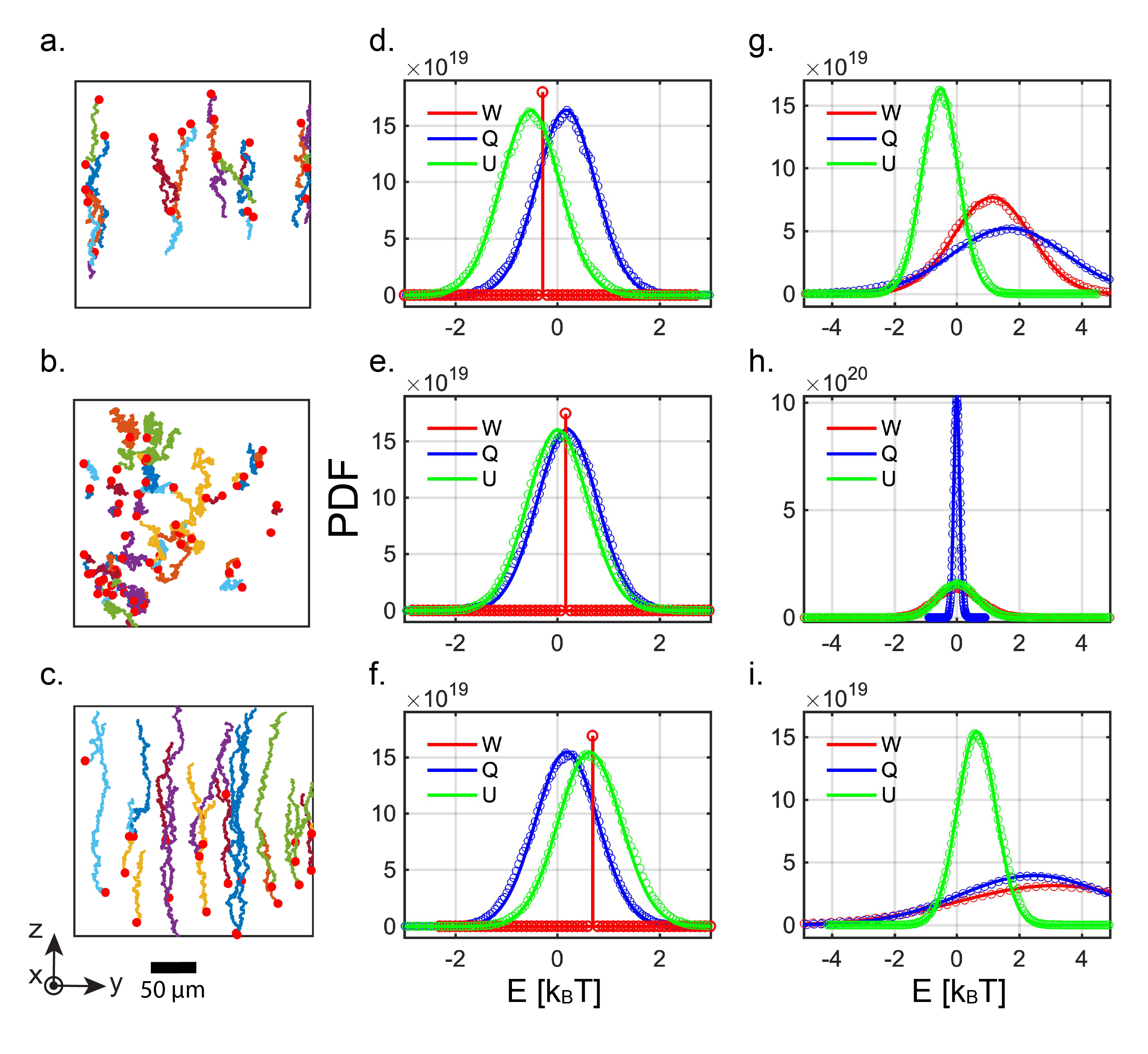}
    \caption{(a)-(c). Experimental trajectories presented in 3 different regimes of external flow induced using laser powers of 16, 24 and 36 mW respectively with the starting point of each trajectory marked with a red spot. The associated energetic PDF of $W$, $Q$ and $\Delta U$ are calculated from these trajectories with a time lag of 1s, and plotted in units of $k_BT$ for the two following cases. Panels (d)-(f) correspond to the real, physical, case where the drift is performing a GT between the laboratory frame and the fluid rest frame. In this case, the PDF of $W$ has the form  of a Dirac distribution showing its deterministic nature. Panels (g)-(i) display the energetic PDF for the case where one considers the drift stemming from an external force field. The differences clearly reveal the ambiguity discussed in the main text. }
\label{fig5}
\end{figure}
%%%%

We finally want to address an ambiguity of Eq. (\ref{PDFLab}) that has far-reaching consequences. The ambiguity stems from the fact that the coarse-grained description does not give the possibility for discriminating the drift term that acts as a GT from an external force field added to the sedimentation one. Indeed, Eq. (\ref{PDFLab}) takes exactly the form of a motional PDF in the presence of a ``resulting'' force field shifting the PDF of the free Brownian motion by $(v_0+v_{\rm sed})\Delta t$. In that case, i.e. if one assigns $\gamma v_0$ to a force, the external force, expressed in the laboratory frame, becomes $F(z,t) = -\nabla U(z,t) + \gamma v_0$. This immediately leads to modified stochastic energetic production rates with $\dot{\tilde{q}}=(\gamma v_0-\Delta \rho Vg)\dot{{z}}$, $\dot{\tilde{w}} =\gamma v_0 \dot{{z}}$, and $\dot{\tilde{u}} =\Delta\rho V g \dot{{z}}$ expressed in the laboratory frame. As clearly seen in this case when $\gamma v_0$ is considered as an external force, the stochastic energetic production rates, evaluated in different inertial frame always with a different $v_0$, are \textit{no longer} GI.

The energetic PDF corresponding to this ``wrong-case'' scenario write as
\begin{eqnarray}
&&\tilde{P}(Q,\Delta t) = \frac{\exp {\left( - \frac{[Q-\gamma (v_0+v_{\rm sed})^2\Delta t]^2}{4D\Delta t  \gamma^2( v_0+v_{\rm sed})^2}\right)}}{\gamma |v_0+v_{\rm sed}|\sqrt{2\pi D \Delta t}} \\
&&\tilde{P}(U,\Delta t) =\frac{\exp{\left( - \frac{ [U-\Delta \rho V g(v_0 + v_{\rm sed})\Delta t]^2}{4D\Delta t  \gamma ^2 v_{\rm sed}^2}\right)}}{\gamma |v_{\rm sed}|\sqrt{2\pi D \Delta t}}  \\
&&\tilde{P}(W,\Delta t) =\frac{\exp{\left( - \frac{[W - \gamma v_0( v_0 + v_{\rm sed})\Delta t ]^2}{4D\Delta t  \gamma ^2 v_0^2}\right)}}{\gamma v_0\sqrt{2\pi D \Delta t}}
\end{eqnarray}
and they are plotted in Fig. \ref{fig5} panels (g-i). Although, these PDF do not violate the first law, the contrast with respect to the frame invariant PDF of Eqs. (13-15) gives a striking illustration of the energetic consequences of assigning to the drift term the role of an external force rather than treating it as induced by a change of reference frame. 

These discrepancies correspond to the fundamental ambiguity in the interpretation of the trajectories measured in the laboratory frame and displayed in Fig. \ref{fig5} (a-c). Indeed, without a previous visualization of the fluid current, it is impossible to anticipate that a change of reference frame is operating on the system. One can thus be led to analyse the modified dynamics of the recorded trajectories as due to a drag force exerted by the illuminating laser on the colloidal ensemble of spheres and acting against sedimentation. The problem resides precisely in the fact that this viewpoint leads, as we just showed, to a totally wrong energetic balance. It remains problematic as long as one models in the laboratory reference frame diffusion and transport by an external force field, without having recognized before hand the comoving frame in which the analysis must be set, with a drift term properly treated in relation with GT. This however is not always possible and can impact dramatically the energetic analysis of force measurements performed within currents, as found for instance in micro- and nanofluidics when studying colloidal transport phenomena from mechanosensitive and mechanoresponsive points of view.   

\section{Conclusion}

Our setup has given us the possibility to induce a controllable, stationary and laminar convection flow inside a fluidic cell by locally heating the fluid (water) under laser illumination. This flow, dragging against the sedimentation a colloidal ensemble dispersed inside the cell, corresponds to a Galilean transform interconnecting the comoving fluid reference frame to the laboratory frame where the Brownian trajectories are recorded.

We have verified experimentally the principle of ``weak'' GI for coarse-grained diffusive systems, and we have derived the expressions for the stochastic energetics production rates and the associated probability density functions that yield a frame invariant formulation of the first law. We emphasized the crucial importance of recognizing in the drift term the signature of a Galilean transform in order not to interpret it as an external force field acting on the colloidal ensemble. We explicitly evaluated the energetic balance in this wrong-case scenario in order to illustrate its strong difference with respect to the frame invariant energetic balance. 

This led us to conclude that misinterpreting the actual role of currents in Brownian experiments eventually leads to violate Galilean relativity that remains central for diffusive systems despite their coarse-grained structure. Our experimental scheme has therefore given us the opportunity to show how mistaking reference frame changes for external force fields leads to nonphysical conclusions. This obviously has practical implications in the context of force measurements in external flows, situations found for instance in soft matter physics and biophysics. 

\section{Acknowledgments}

This work was supported by the French National Research Agency (ANR) through the Programme d'Investissement d'Avenir under contract ANR-17-EURE-0024, the ANR Equipex Union (ANR-10-EQPX-52-01), the Labex NIE (ANR-11-LABX-0058 NIE) and CSC (ANR-10-LABX-0026 CSC) projects, the University of Strasbourg Institute for Advanced Study (USIAS) (ANR-10-IDEX-0002-02).

\appendix

\section{Appendix}

\section{A. Experimental setup and sample preparation}
\label{AppExp}

Our experiment consists in illuminating with horizontally two counterpropagating laser beams a colloidal dispersion of micron-sized melamine spheres, diffusing and sedimenting inside a cuvette filled with water. The same setup has already been described and exploited in a weak force measurement context in \cite{LiPRA2019}. With balanced power in the counterpropagating beams, the laser induces convection within the fluid along the vertical $z-$direction with no radiation pressure effect at play. This perfect cancellation of radiation pressure makes the diffusion dynamics along the $y-$axis look like a free normal Brownian motion. Along the $z-$axis, the Brownian motion is performed within a laminar flow. This diffusion dynamics can be analysed by looking at real-time colloidal trajectories using the optical setup shown in Fig. \ref{figS1} and recorded by tracking the successive positions of the particles using an algorithm adapted from \cite{Trackmate}. 

Our setup has important features. First, the cuvette has large dimensions compared to the size of the imaging region-of-interest (ROI) that only extends over a small central region far away from all walls. This, together with the small-volume fraction of the colloidal dispersion allows us to neglect the influence of possible boundary-wall and inter-particle-interaction effects on the diffusion dynamics. Also, this far-from-walls feature is a key point for the laminar flow generated by collective sedimentation and the laser induced convection detailed in Appendix B. Second, the illumination conditions are set so that within the imaging ROI, the Gaussian profile of the laser beam is uniform along the horizontal optical $y-$axis considering that the Rayleigh length is set much larger than the width of the cuvette. Since the waist of laser is much larger than the diameter of a colloidal sphere, the close to plane-wave illumination conditions minimize any gradient contribution to the optical force field, yielding no external force along the $z-$axis except gravity. 

The samples are prepared from an initial dispersion (2.5$\%$ mass-volume ratio) of melamine microparticles of diameter $d = 0.94 \pm 0.05\mu $m purchased from microParticle GmbH, weakly doped with a fluorescent dye for a most efficient detection in water. We dilute the dispersion $\sim 10^4$ times with ultra-pure water and fill the cuvette with the colloidal dispersion to ca. $10^{-6}$ low-volume fraction. The filled cuvette is covered and sealed with a vacuum grease to prevent water evaporation and to isolate the fluid from other environmental influence. In addition, the cuvette and its cover are exposed at least 1 h to UV light to ensure the absence of any bacterial contaminant. The sample is also grounded to remove any electrostatic charge on the surface of the cuvette which generate a resulting external force on particles. Before performing our experiments, we leave for about 1 h the sample relaxing in its holder until well thermalised with the environment. This also ensures that potential colloidal aggregates have sedimented at the bottom of the cuvette. The temperature of the laboratory is controlled with a thermal precision better than $1 K$ at room temperature.

These illumination conditions, together with the great care of the sample preparation, allows us to exclude, as much as it is possible, all potential perturbing influences on the colloidal diffusion dynamics. This ensures us the capacity to quantitatively analyse the diffusion dynamics along the vertical $z$ direction. 

\begin{figure}[htb]
  \centering{
    \includegraphics[width=1\linewidth]{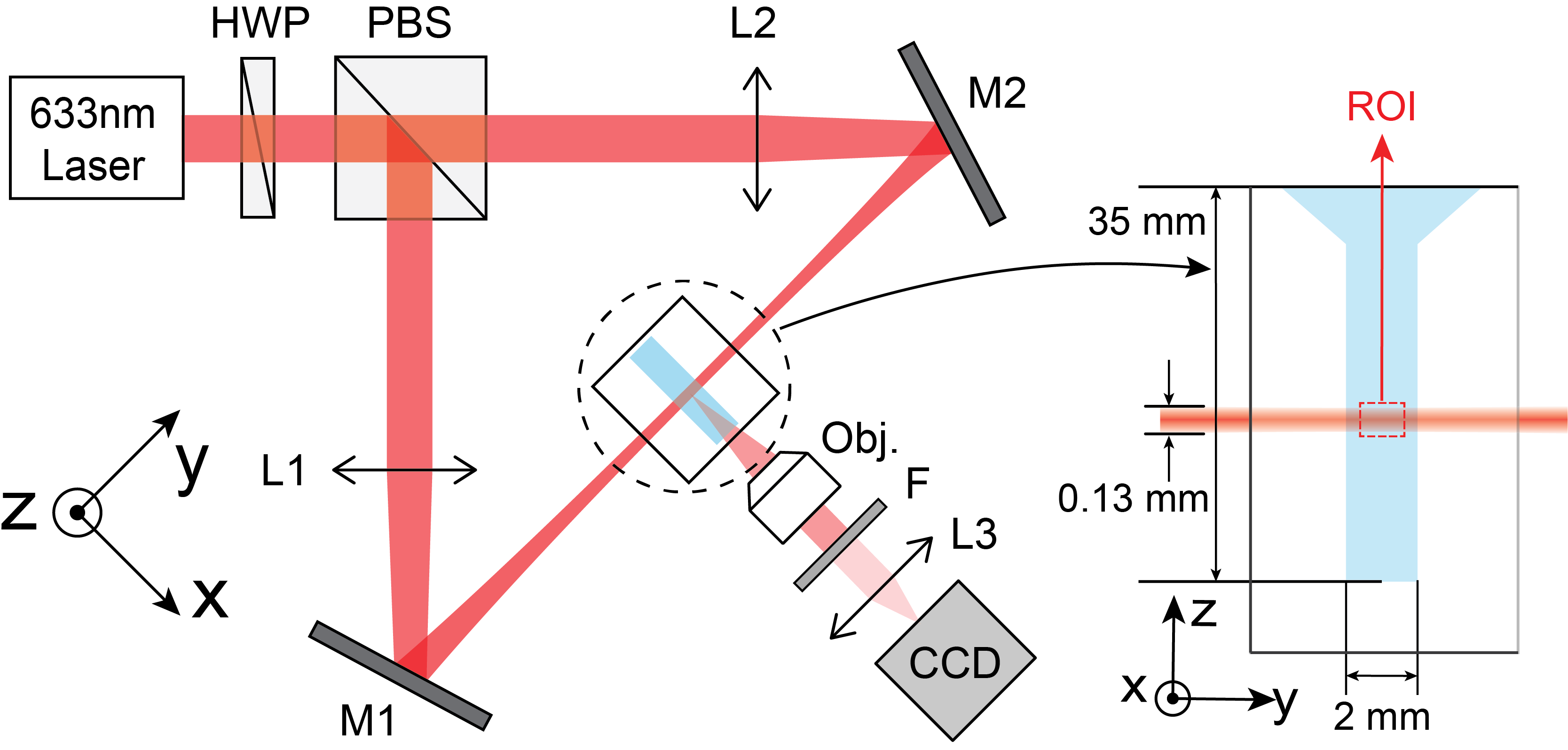}}
    \caption{Schematic of the optical experimental setup already detailed in \cite{LiPRA2019}. Using a halfwave plate (HWP), a polarizing beam-splitter (PBS) and two mirrors (M1, M2), a linearly polarized single-mode laser beam (wavelength $633$ nm, $200$ mW, TEM$_{00}$) can be split into two noninterfering (cross-polarized) counterpropagating beams of identical intensity (using the HWP for the fine intensity balance). A microscope objective (NA = 0.25, 20$\times$) collects the fluorescence of dye-doped melamine spheres (diameter d = 0.94 $\pm 0.05 \mu$m, from microParticle GmbH) diffusing in water inside the cell with the help of a filter F. The particles are imaged on a CCD camera at a frame rate f = 120 Hz. The cell consists of a quartz cuvette of dimensions $10(x) \times 2(y) \times 35(z)$ mm$^3$ chosen such that the imaging region-of-interest (ROI) is located far away from any wall, allowing us to neglect safely any perturbation of the walls on the diffusion dynamics.}
\label{figS1}
\end{figure}

%%%%
\section{B. Laser induced convection}
\label{AppCnvt}

The temperature and velocity profiles of the laser-induced convection flow within our experimental ROI --see Figs. \ref{fig1} and \ref{figS1}-- can be calculated analytically by exploiting the symmetries of our system and its illumination conditions that enable to decouple the two heat transfer and Navier-Stokes (NS) equations. 

\begin{figure}[htb]
  \centering{}
    \includegraphics[width=0.5\linewidth]{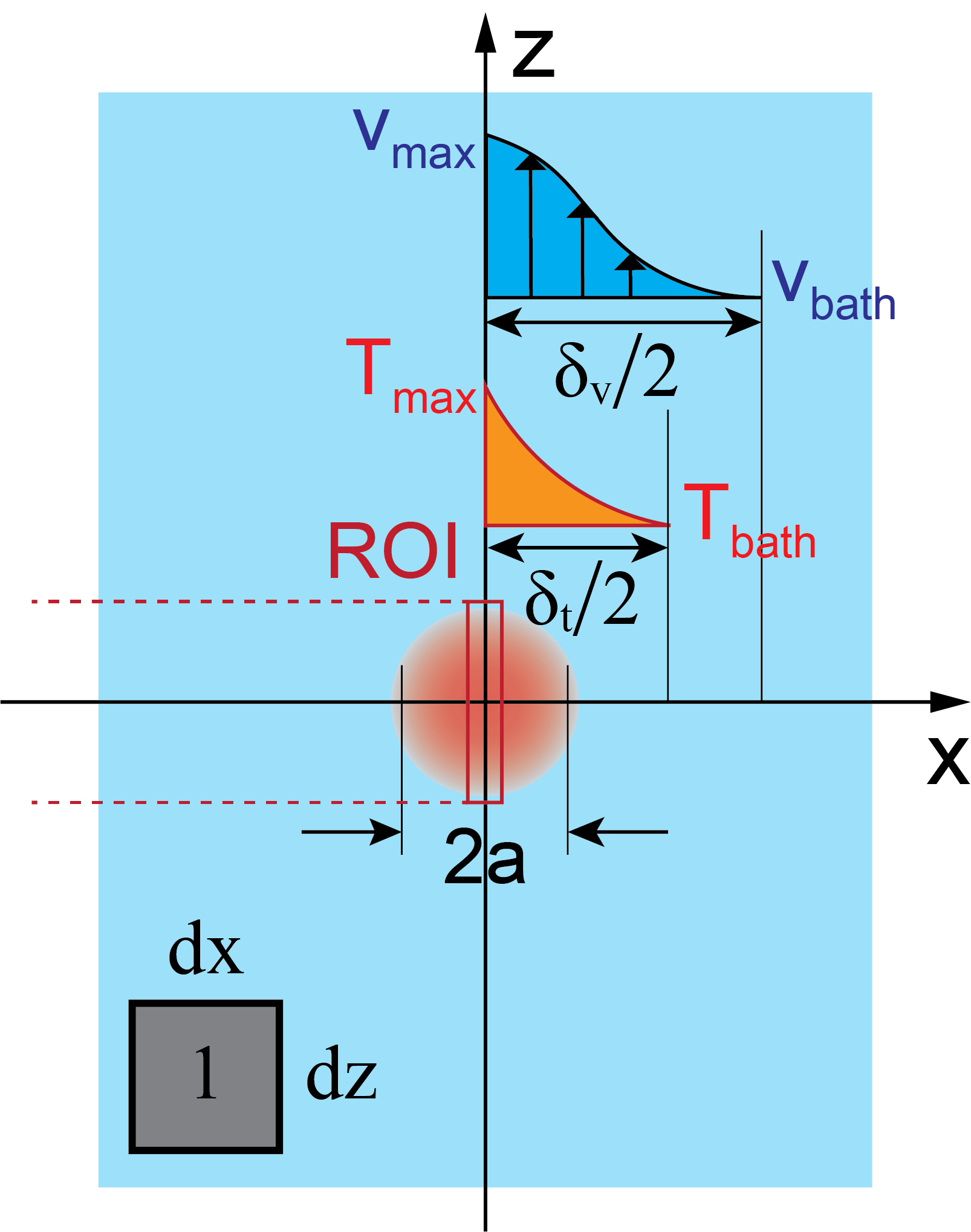}
    \caption{Scheme of the  coordinate system used for estimating the laser induced convection velocity profile. As the system under consideration has a symmetry along the optical axis, the problem is solved within the $(x,z)$ plane. Therefore, the elementary volume is set as $dx\cdot 1 \cdot dz$. The waist of the laser projected in the $(x,z)$ plane is $w_0=\sqrt{2}a$. The displayed temperature boundary layer thickness $\delta_T$ and the velocity boundary layer thickness $\delta_v$ correspond to the region beyond which the quantity (temperature or velocity) decays to that of the unperturbed bath. }
\label{figS2}
\end{figure}

With the coordinate setting shown in Fig. \ref{figS2}, we first look at the heat transport Eq. (\ref{Eqcoup1}) in the steady state ($\partial_t T = 0$): 
\begin{equation}
(\bm{v}\cdot \nabla)\delta T - \frac{k}{c_p\rho} \nabla^2 \delta T = \frac{\dot q_L}{c_p\rho}.
\end{equation}
As discussed in the main text, the convection flow inside the ROI is essentially induced along the $z-$axis and measured to reach velocities of the order of $10^{-6}$ m/s. Considering that a laser of ca. $100$ mW will lead to an ca. mK increase in the local temperature on a typical scale of the order of the waist (in our case, ca. $100~\mu$m), a simple scaling argument leads to a convective contribution of the Lagrangian derivative $(\bm v\cdot \nabla) \delta T\sim 10^{-5}$ much smaller than the contribution associated with the volumetric heat rate $\dot{q}/(\rho c_p)\sim 10^{-1}$ K/s at such laser powers. This scaling simplifies the heat equation to: 
\begin{equation}
 \nabla^2 \delta T = -\frac{\dot q_L}{k}
\label{eqF1}
\end{equation}
with $\dot{q}_L = A P_0/\pi a^2 \exp{(-r^2/a^2)}$ and $r^2 = x^2+z^2$. In such polar coordinates, Eq. (\ref{eqF1}) can be solved through the following steps, starting with:
\begin{equation}
\frac{1}{r} \frac{\partial}{\partial r}\left(r\frac{\partial}{\partial r}\delta T\right) = -\frac{AP_0}{\pi a^2 k}\exp\left(-\frac{r^2}{a^2}\right)
\end{equation}
first integrated into 
\begin{eqnarray}
r\frac{\partial}{\partial r}\delta T &=& - \frac{AP_0}{\pi a^2 k} \int\limits_{0}^{r} \exp \left(-\frac{u^2}{a^2}\right) u du  \nonumber \\
&=&- \frac{AP_0}{2 \pi k} \left(  1 - \exp \left(- r^2/a^2\right)\right)
\end{eqnarray}
with the limit condition $r \partial_r \delta T|_{r = 0} = 0$. By reminding that $\ln '(r)-E_1'(r)= 1/r - \exp (-r) / r$ with $E_1$ the exponential integral function, we can further integrate from $r$ to the temperature boundary layer $\delta_T / 2$ defined as the radial distance at which $\delta T(\delta_T / 2) = 0$, giving
\begin{eqnarray}\nonumber
\delta T(r) &=& - \frac{AP_0}{2 \pi k} \left(\ln (r) + \frac{1}{2} E_1\left( \frac{r^2}{a^2} \right) \right. \\
&& \left. -\ln (\delta_T/2) - \frac{1}{2} E_1\left( \frac{\delta_T^2}{4a^2} \right) \right).
\label{solT}
\end{eqnarray}

The temperature profile is then injected into the NS equation Eq. (\ref{Eqcoup2}) within the Boussinesq's approximation and the condition of incompressibility in the vicinity of our ROI shown in Fig. \ref{figS2} that yield the convection velocity field ${\bf v}({\bf r})=v_0(x,y)\hat{z}$ and cancel the convective contribution $({\bf v}\cdot \nabla){\bf v}=0$ to the Lagrangian derivative. With ${\bf g}=-g\hat{z}$, the NS equation becomes in the steady-state:
\begin{equation}
\nu\left(\frac{\partial^2 v_0}{\partial x^2} + \frac{\partial^2 v_0}{\partial y^2} \right) = -g\alpha\delta T(r) + \frac{\phi\Delta \rho g}{\rho}.
\label{eqNS1}
\end{equation}
where $\nu=\mu/\rho$ the kinematic viscosity. We will further exploit the consequence of incompressibility with $\partial_z v_0 = 0$ in order to solve Eq. (\ref{eqNS1}) at $z \sim 0$ where we can simplify the problem to a one-dimensional one with $\delta T(r = \sqrt{x^2 + z^2}) \sim \delta T(x)$. The translational invariance (within the ROI) of $\delta T(x)$ along the $y-$axis allows us to apply a separation of variable
\begin{equation}
v_0 (x,y) = u_z(x) + w_z(y)
\end{equation}
that consists in decomposing convection into two drives: one thermal with the source term $-g\alpha\delta T(x)$ that only depends on $x$ and a second associated with the collective body force ${\phi\Delta \rho g/\rho}$ determining for the fluid the $y-$dependence of the convection velocity according to
\begin{eqnarray}
\frac{\partial^2 u_z (x)}{\partial x^2} &=& -\frac{g\alpha}{\nu} \delta T(x)  \\
\frac{\partial^2 w_z (y)}{\partial y^2} &=& \frac{\phi\Delta \rho g}{\mu}
\end{eqnarray}

Applying on $w_z$ a boundary condition of the first type
\begin{equation}
\frac{\partial w_z }{\partial y}\bigg|_{y=0} = 0
\end{equation}
where thus $w_z(y=0)=v_{\rm bath}$ is extremal, and a boundary condition of the second type
\begin{equation}
w_z (\pm l_y)=0,
\end{equation}
where $2l_y$ corresponds to the length of the cuvette along $y$ ($l_y = 1$mm). With these, the solution reads directly as:
\begin{equation}
w_z(y) = \frac{\phi\Delta \rho g}{2\mu} (y^2 - l_y^2)
\label{solwz}
\end{equation}

Considering that water has a Prandtl number ${\rm Pr}$ larger than one, the velocity boundary layer thickness is larger than the temperature boundary layer thickness according to $\delta_v = \sqrt{{\rm Pr}}\delta_T > \delta_T$. Keeping in mind  that $\delta T(x)$ is non-zero only within $\delta_T$, the $u_z(x)$ solution is piecewise, defined on the matching intervals:
\begin{equation}
u_z(x) = 
\left\{
	\begin{aligned}
	&u_2(x)&, x \in [-\frac{\delta_v}{2}, -\frac{\delta_T}{2})\\
	&u_1(x)&, x \in [\frac{\delta_T}{2},\frac{\delta_T}{2}]\\
	&u_3(x)&, x \in (\frac{\delta_T}{2}, \frac{\delta_v}{2}] ,
	\end{aligned}
\right.
\end{equation}
with a piecewise differential equation:
\begin{eqnarray}\nonumber
&&\frac{\partial^2 u_1}{\partial x^2} = -\frac{g\alpha A P_0}{4 \pi k\nu} \left( 2\ln \left( \frac{2x}{\delta_T} \right) + E_1\left( \frac{x^2}{a^2} \right) - E_1\left( \frac{\delta_T^2}{4a^2} \right) \right)   \\
&&\frac{\partial^2 u_{2,3}}{\partial x^2} = 0
\end{eqnarray}
with first type boundary conditions
\begin{eqnarray}
&& u_1(-\delta_T/2) = u_2(-\delta_T/2) \label{BC11} \\
&& u_1(\delta_T/2) = u_3(\delta_T/2) \label{BC12} \\
&& u_2(-\delta_v/2) = u_3(\delta_v/2) = w_z(y=0) = v_{\rm bath} \label{BC13}
\end{eqnarray}
and second type boundary conditions
\begin{eqnarray}
&& \frac{\partial u_1}{\partial x}\bigg|_{x=0} = 0 \label{BC21}\\
&& \frac{\partial u_1}{\partial x}\bigg|_{x=-\delta_T/2} = \frac{\partial u_2}{\partial x}\bigg|_{x=-\delta_T/2} \label{BC22}\\
&& \frac{\partial u_1}{\partial x}\bigg|_{x=\delta_T/2} = \frac{\partial u_3}{\partial x}\bigg|_{x=\delta_T/2} \label{BC23}
\end{eqnarray}
Defining
\begin{equation}
K = \frac{g\alpha A P_0}{4 \pi k\nu}, ~ C_0 = - 2\ln \left( \frac{\delta_T}{2} \right) - E_1\left( \frac{\delta_T^2}{4a^2} \right),
\end{equation}
simplifies the equation to be solved into:
\begin{eqnarray}\nonumber
&&\frac{\partial^2 u_1}{\partial x^2} = K\left(2\ln x + E_1 \left(\frac{x^2}{a^2} \right) + C_0 \right)  \\
&&\frac{\partial^2 u_{2,3}}{\partial x^2} = 0.
\end{eqnarray}
The first integration gives
\begin{equation}
\begin{aligned}
\frac{\partial u_1}{\partial x} = &K \bigg( 2x \ln(x) -2x +C_0 x + x E_1\left( \frac{x^2}{a^2}\right) \\
&+ a \sqrt{\pi} {\rm erf}\left(  \frac{x}{a} \right) \bigg) +C_1
\label{dxu1}
\end{aligned}
\end{equation}
with the error function ${\rm erf}(x)$ defined as
\begin{equation}
{\rm erf}(x) = \frac{2}{\sqrt{\pi}} \int_0^{x} e^{-w^2} dw.
\end{equation}
Using the second type boundary condition of Eq. (\ref{BC21}) determines $C_1 = 0$.
For $u_2$ and $u_3$, we have,
\begin{eqnarray}
&&\frac{\partial u_2}{\partial x} = C_1^{\prime} \\
&&\frac{\partial u_3}{\partial x} = C_1^{\prime\prime}.
\end{eqnarray}
Using the additional second type boundary conditions of Eqs. (\ref{BC22},\ref{BC23}), fixes
\begin{eqnarray}
&&C_1^{\prime} = K\left(\delta_T - a\sqrt{\pi} {\rm erf}\left( \frac{\delta_T}{2a} \right) \right)  \\
&&C_1^{\prime\prime} = - K\left(\delta_T - a\sqrt{\pi} {\rm erf}\left( \frac{\delta_T}{2a} \right) \right) = -C_1^{\prime}.
\end{eqnarray}
Further integrating Eq. (\ref{dxu1}) gives:
\begin{equation}
\begin{aligned}
u_1 =& K \bigg( x^2 \ln (x) - \frac{x^2}{2} - x^2 + \frac{C_0}{2} x^2 + \int x E_1\left( \frac{x^2}{a^2}\right) dx \\
&+ a^2 \sqrt{\pi} \int {\rm erf} \left(  \frac{x}{a} \right) d \frac{x}{a} \bigg) + C_2 .
\end{aligned}
\end{equation}
where
\begin{equation}
 \int {\rm erf} \left(  \frac{x}{a} \right) d \frac{x}{a} = \frac{x}{a} {\rm erf} \left(  \frac{x}{a} \right) + \frac{1}{\sqrt{\pi}} e^{-x^2/a^2} 
\end{equation}
and
\begin{equation}
\int x E_1\left( \frac{x^2}{a^2}\right) dx = \frac{a^2}{2} \left( \frac{x^2}{a^2} E_1( \frac{x^2}{a^2}) - e^{- {x^2}/{a^2}} \right).
\end{equation}
This yields the final solutions that read for $u_1$ as:
\begin{equation}
\begin{aligned}
u_1 =& K \bigg[ x^2 \ln (x) + \frac{C_0 - 3}{2} x^2 \\
&+ a^2 \sqrt{\pi} \left( \frac{x}{a} {\rm erf} \left(  \frac{x}{a} \right) + \frac{1}{\sqrt{\pi}} e^{-x^2/a^2} \right) \\
&+ \frac{a^2}{2} \left( \frac{x^2}{a^2} E_1\left( \frac{x^2}{a^2} \right) - e^{-x^2/a^2} \right) \bigg] + C_2 
\end{aligned}
\end{equation}
and for $u_2$ and $u_3$ as:
\begin{eqnarray}
&&u_2 = C_1^{\prime}x + C_2^{\prime}  \\
&&u_3 = - C_1^{\prime}x + C_2^{\prime\prime}.
\end{eqnarray}
The constants are now determined by using the first type of boundary conditions:
\begin{eqnarray}
C_2^{\prime} &=& C_2^{\prime\prime} = v_{\rm bath} + \frac{\delta_v}{2}C_1^{\prime}   \\
C_2 &=& K\left( -\frac{1}{8}\delta_T^2 + \frac{\delta_T \delta_v}{2} - \frac{a^2}{2} e^{-\delta_T^2/4a^2} \right.  \nonumber  \\
&&\left.  - \sqrt{\pi} \frac{a\delta_v}{2} {\rm erf} \left( \frac{\delta_T}{2a} \right) \right) + v_{\rm bath}.
\end{eqnarray}

These solutions determine the complete convection velocity $v_0 (x,y)=u_z(x)+w_z(y)$ within the velocity boundary layer thickness $\delta_v$. Accordingly, the velocity measured inside the ROI corresponds to the velocity evaluated at $(x=0,y=0)$: 
\begin{equation}
\begin{aligned}
v_0(0,0) =& \frac{g\alpha A P_0}{4 \pi k\nu} \bigg[ \frac{a^2}{2} -\frac{1}{8}\delta_T^2 + \frac{\delta_T \delta_v}{2} - \frac{a^2}{2} e^{-\delta_T^2/4a^2} \\
& -\sqrt{\pi} \frac{a\delta_v}{2} {\rm erf} \left( \frac{\delta_T}{2a} \right) \bigg]  + v_{\rm bath}.
\end{aligned}
\end{equation}

Evaluating this velocity and its evolution with the laser power $P_0$ demands to determine the boundary layer thicknesses $\delta_T$ and $\delta_v$ together with $v_{bath}$. $v_{bath}$ is evaluated by solving the NS equation in the absence of any laser heating, under the sole influence of collective sedimentation. The boundary layer thicknesses can be determined through a simple scale analysis, where according to \cite{BejanBook}: 
\begin{equation}
\delta_T = {\rm Ra}^{-1/4}L, 
\label{delt}
\end{equation}
taking for the characteristic length $L\sim 2\times w_0$ and with ${\rm Ra}$ the Rayleigh number defined by
\begin{equation}
{\rm Ra} = \frac{g\alpha c_p \rho \delta T_{\rm max} L^3}{k\nu}.
\end{equation}
For a medium with large Prandtl number ${\rm Pr} > 1$ just like water is:
\begin{equation}
\delta_v = \sqrt{{\rm Pr}}\delta_T.
\label{delv}
\end{equation}
According to Eq. (\ref{solT}), $\delta T_{\rm max} = \delta T(0)$. Using the known identity for the exponential integral function
\begin{equation}
E_1(x) = -\gamma - \ln(x) - \sum_{n=1}^{\infty} \frac{(-x)^n}{n\cdot n!}
\end{equation}
where $\gamma$ is the Euler-Mascheroni constant, we expand $E_1(x)$ in the $x\rightarrow 0$ limit as:
\begin{equation}
\lim_{x\rightarrow0} E_1(x) = -\gamma - \ln(x).
\end{equation} 
Therefore, $\delta T_{\rm max}$ reads
\begin{equation}
\delta T_{max} =  \frac{AP_0}{4 \pi k} \left( 2 \ln \left( \frac{\delta_T}{2a} \right) + \gamma + E_1\left( \frac{\delta_T^2}{4a^2} \right) \right)
\label{Tmax}
\end{equation} 
In the limit where $\displaystyle \frac{\delta_T^2}{4a^2} \gg 1$, $E_1\left( \displaystyle \frac{\delta_T^2}{4a^2} \right) \sim 0$, so that by substituting the expression of $\delta_T$ -cf. Eq. (\ref{delt})- into Eq. (\ref{Tmax}) yields an implicit equation for determining $\delta T_{\rm max}$ as:
\begin{equation}
\delta T_{\rm max} - \frac{AP_0}{4 \pi k} \left[ 2\ln\left( \frac{1}{2} \left(\frac{g\alpha c_p \rho \delta T_{\rm max} L^3}{k \nu} \right)^{-1/4} \frac{L}{a} \right) + \gamma \right] = 0 .
\end{equation}
that is solved and which solution $\delta T_{max}$ is used for determining $\delta_T$ according to Eq. (\ref{delt}) and $\delta_v$ through Eq. (\ref{delv}). 
Finally, the velocity at the centre of the ROI, i.e. at the centre of the cuvette is
\begin{equation}
\begin{aligned}
v_0(0) =& \frac{g\alpha A P_0}{4 \pi k\nu} \bigg[ \frac{a^2}{2} -\frac{1}{8}\delta_T^2 + \frac{\delta_T \delta_v}{2} - \frac{a^2}{2} e^{-\delta_T^2/4a^2} \\
& -\sqrt{\pi} \frac{a\delta_v}{2} {\rm erf} \left( \frac{\delta_T}{2a} \right) \bigg]  - \frac{\phi \Delta\rho g}{2\mu} l_y^2 .
\label{vmax}
\end{aligned}
\end{equation}
This is the velocity $v_0$ engaged in the GT that connects the two lab $S$ and comoving $S^\prime$ reference frames. The particle mean velocity noted $v_z$ in the main text, can be estimated by adding to the convection flow velocity $v_0$ the single particle sedimentation velocity $v_{\rm sed}$ according to $v_z=v_0+v_{\rm sed}$. The profile of this convection velocity through the ROI is displayed in Fig. \ref{fig2} in the main text, using values that correspond to our experimental conditions: $\alpha = 0.0003$ K$^{-1}$, $A = 0.3$ m$^{-1}$, $k = 0.61$ W$\cdot$K$^{-1}\cdot$m$^{-1}$, $\nu = 8.72\times 10^{-7}$ m$^2/$s, $c_p = 4.18\times10^3 $J$\cdot$kg$^{-1}\cdot$K$^{-1}$, $\Delta \rho = 513$ kg$/$m$^3$, $\mu = 0.87 \times 10^{-3} $N$\cdot$s$\cdot$m$^{-2}$. The boundary layer thicknesses $\delta_T$ and $\delta_v$ take the smallest value between the ones obtained by Eq.(\ref{delt}) and (\ref{delv}) and the minimum length found in the cuvette $l_y$ (= 2mm). The mean volume fraction $\phi$ can be estimated by comparing the experimental data in Fig. \ref{fig4} (d). The value of $\phi$ is ca. $6.8 \times 10^{-7}$ is in good agreement with the expected value ($\phi\sim10^{-6}$) and the evolution of the observed velocity combining convection ($v_0$) and single sedimentation velocity ($v_{\rm sed}$) with the illuminating laser power $P_0$ is displayed in Fig \ref{figS3}. Negative values at low laser powers correspond to the situation where the laser-induced convection flow is not strong enough to compensate for the collective sedimenting effect of the body force acting on the colloidal dispersion within the ROI.

\begin{figure}[htb]
  \centering{}
    \includegraphics[width=0.75\linewidth]{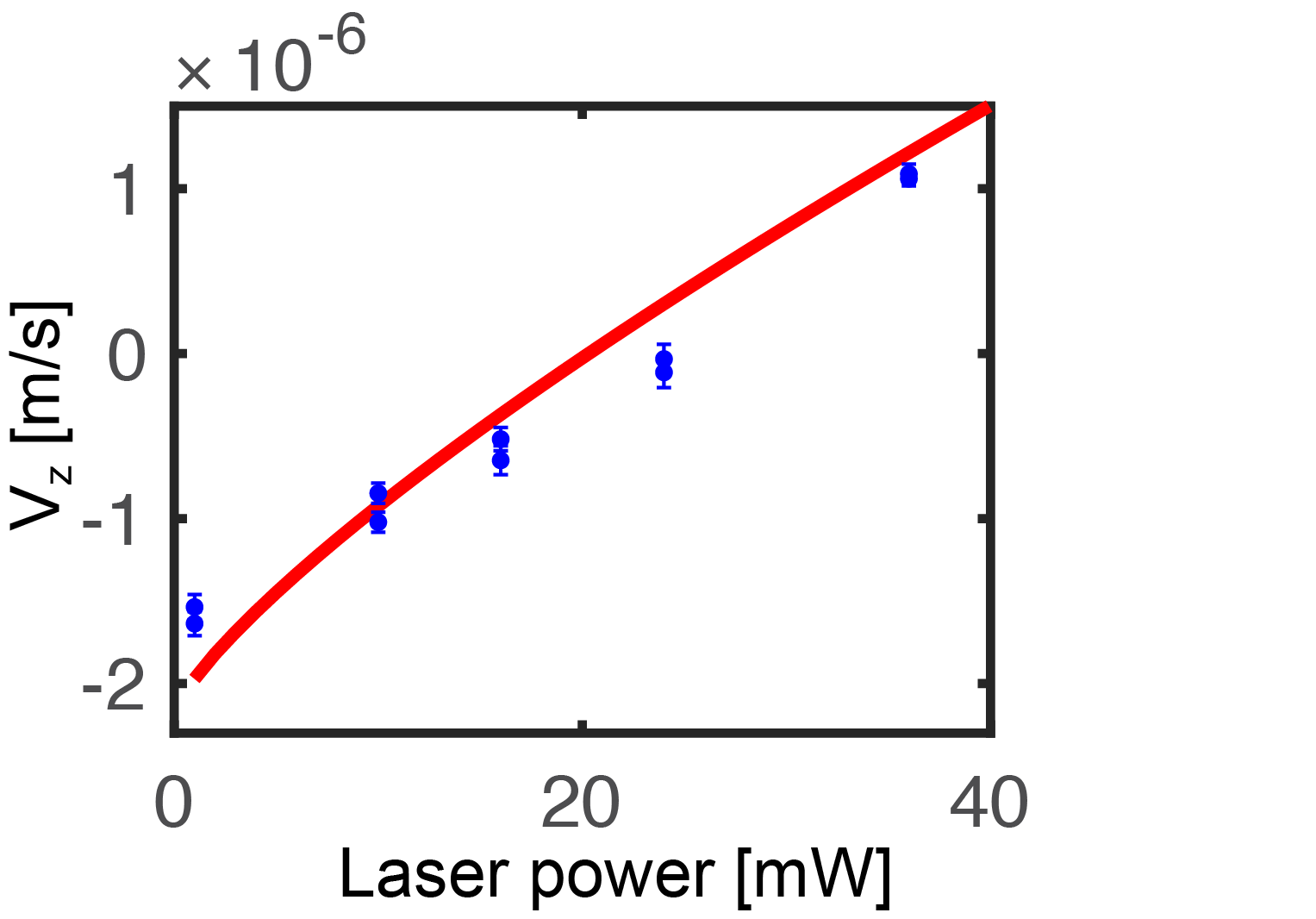}
    \caption{Comparison between the evolution of the particle velocity $v_z$ evaluated by Eq. (\ref{vmax}) with the illuminating laser power $P_0$ --using $\phi = 6.8 \times 10^{-7}$-- (red line) and values of the mean particle velocity measured experimentally (blue dots).}
\label{figS3}
\end{figure}

%%%%
\section{C. Stochastic entropy under Galilean transformation}
\label{AppEtpy}

Following \cite{SeifertPRL2005}, the stochastic entropy for single particle $s_{tot}$ can be split into an entropy associated with the single particle trajectory (particle configuration) $s_p$ and an entropy $s_m$ associated with heat dissipated into the thermal bath.Both are initially defined as:
\begin{equation}
s_p = -k_B\ln(P(\bm x,t))
\end{equation}
\begin{equation}
s_m = \frac{q}{T}
\end{equation}
where $P(\bm x,t)$ is the probability density function of position for the single particle motion, and $q$ is the dissipated heat from the particle to the thermal bath, with $T$ the temperature of this surrounding bath. This way, the total entropy production rate can be written as:
\begin{equation}
\dot s_{tot} = \dot s_p + \dot s_m=  -k_B\frac{\partial_t P(\bm x,t)}{P(\bm x,t)} - k_B\frac{\partial_x P(\bm x,t)}{P(\bm x,t)} \dot{\bm x} + \frac{\dot q}{T}
\end{equation}
The entropy production rates in the two different inertial reference frames $S$ and $S^\prime$ are related by a GT $\bm x^{\prime} = \bm x - \bm v_0t$
can be calculated for each entropy part. For the trajectory-dependent entropy production rate, the results demonstrated in the main text related to the ``weak'' Galilean Invariance (GI) for the probability density function (PDF) in the different inertial reference frames --see Eq.(\ref{Ptrans})-- lead to an entropy in the co-moving reference frame $s_p^{\prime}$ that can be written as:
\begin{equation}
s_p^{\prime} = -k_B\ln(P^{\prime}({\bm x}^{\prime},t)) = -k_B\ln(P({\bm x}^{\prime} + \bm v_0 t,t))
\end{equation}
For the production rate of that trajectory-dependent entropy, we easily demonstrate the GI:
\begin{equation}
\begin{aligned}
\dot{{s}}^{\prime}_p & = -k_B \frac{\partial_t P({\bm x}^{\prime} + \bm v_0 t,t)}{P({\bm x}^{\prime} + \bm v_0 t,t)} - k_B\frac{\partial_x P({\bm x}^{\prime} + \bm v_0 t,t)}{P({\bm x}^{\prime} + \bm v_0 t,t)}(\dot{{\bm x}^{\prime}} + \bm v_0)\\
& = -k_B \frac{\partial_t P(\bm x,t)}{P(\bm x,t)} - k_B \frac{\partial_x P(\bm x,t)}{P(\bm x,t)} \dot{\bm x} = \dot s_p
\end{aligned}
\end{equation}
As for the entropy production rate due to heat dissipation, considering that we have already demonstrated that the heat production is the same in the different inertial reference frame, we can easily obtain $\dot s_m = \dot {s'}_m$.  Therefore, combining both rates, we directly arrive to the conclusion that the total stochastic entropy production is Galilean Invariant. 

\bibliography{biblio_galilee}

\end{document}